\documentclass{article}
\usepackage{spconf,amsmath,graphicx}
\usepackage{times}
\usepackage{epsfig}
\usepackage{graphicx}
\usepackage{amssymb}
\usepackage{tabularx}

\usepackage{subcaption}
\usepackage{color}
\usepackage{url}

\title{Learning An Inverse Tone Mapping Network with A Generative Adversarial Regularizer}
\name{Shiyu Ning$^1$, Hongteng Xu$^{2,3}$, Li Song$^1$, Rong Xie$^1$, Wenjun Zhang$^1$}
\address{\small $^1$School of Electronic Information and Electrical Engineering, Shanghai Jiao Tong University\\ 
\small $^2$Department of Electrical and Computer Engineering, Duke University,  $^3$InfiniaML Inc.}
\begin{document}
%
\maketitle
\begin{abstract}
Transferring a low-dynamic-range (LDR) image to a high-dynamic-range (HDR) image, which is the so-called inverse tone mapping (iTM), is an important imaging technique to improve visual effects of imaging devices. 
In this paper, we propose a novel deep learning-based iTM method, which learns an inverse tone mapping network with a generative adversarial regularizer. 
In the framework of alternating optimization, we learn a U-Net-based HDR image generator to transfer input LDR images to HDR ones, and a simple CNN-based discriminator to classify the real HDR images and the generated ones. 
Specifically, when learning the generator we consider the content-related loss and the generative adversarial regularizer jointly to improve the stability and the robustness of the generated HDR images. 
Using the learned generator as the proposed inverse tone mapping network, we achieve superior iTM results to the state-of-the-art methods consistently. 
\end{abstract}

\section{Introduction}
With the development of ultra-high-definition television techniques, the demands on high-dynamic-range (HDR) contents are increasing rapidly these years. 
These demands require us to transfer a large amount of existing low-dynamic-range (LDR) images and videos to HDR contents efficiently and effectively. 
In such a situation, manually transferring is infeasible because of the huge size of the contents, and we need to apply inverse tone mapping (iTM) techniques.

An ideal iTM method should be multi-scale invariant and adaptive to various objects in different conditions, which requires an nonlinear mapping with high complexity. 
However, most of existing methods~\cite{xu2014generalized,huo2014physiological,kovaleski2014high}  simplify the mapping of luminance and that of color as segmented mapping empirically based on histogram equalization and spatial filtering.  
These methods either ignore the nonlinear nature of iTM or the color correlation between the iTMs across different channels. 
Recent research~\cite{Endo2017Deep} apply CNN to generate multi-exposed images and then merge them using traditional merging method.
They mostly focus on dynamic range but not on color gamut, and merging methods are not robust enough.

To solve the challenges mentioned above, we propose a novel inverse tone mapping network (iTMN) based on generative adversarial network (GAN)~\cite{goodfellow2014generative}. 
As shown in Fig.~\ref{fig:1}, we aim to train a U-Net-based HDR image generator~\cite{ronneberger2015u} as our inverse tone mapping network, which transfer LDR images to HDR ones. 
In each iteration, given existing generator the discriminator is updated to distinguish the generated HDR images from the ground truth with higher accuracy. 
Given updated discriminator, we further update the generator by minimizing a content-based loss with the adversarial regularizer related to the discriminator. 
As a result, each generated HDR image is close to the ground truth and the distribution of all generated HDR images is identical to that of training HDR images. 
Applying this learning method, we obtain an state-of-the-art inverse tone mapping network. 

\section{Related Work}
\subsection{Inverse Tone Mapping} 
Inverse tone mapping has been studied for a long time. 
The work in~\cite{akyuz2007hdr} presented a simple linear expansion method to prove that HDR contents could be produced from LDR images without sophisticated process.
A physiological iTM method with low complexity is proposed in~\cite{huo2014physiological}, which is still sensitive to the choice of parameters. 
A generalized histogram equalization method is proposed in~\cite{xu2012no,xu2014generalized}, which shows potentials to generate HDR images. 
Filtering-based methods are also applied to enhance the dynamic range and the details of image~\cite{kovaleski2014high,xu2013single}.
Recently, HDR images are merged by inferred bracketed images in different exposure, which is generated by CNN-based up/down-exposure model~\cite{Endo2017Deep}.
Although these methods achieve the state-of-the-art performance, they still suffer to over-exposure enhancement and color over/under-saturation.
Moreover, none of them apply generative adversarial networks in their models and algorithms.

\begin{figure}[t]
\centering
\includegraphics[width=1\linewidth]{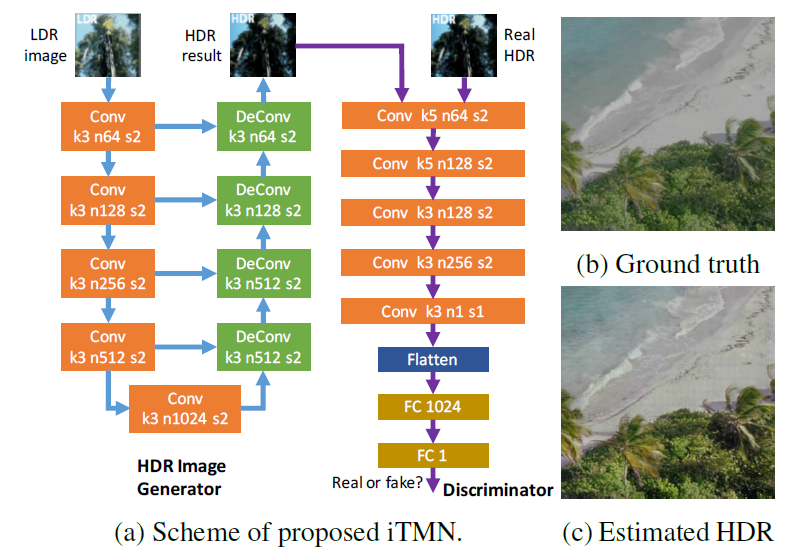}
\vspace{-7pt}
\caption{\small (a) The illustration of our generator and discriminator with specific kernel size (k), number of feature maps (n) and stride (s) for each convolution/deconvolution layer (Conv/DeConv) and number of nodes for each fully-connected layer (FC). (b, c) A typical comparison between the ground truth and our HDR result.}
\label{fig:1}
\end{figure}

\subsection{Deep learning and Image Processing} 
Deep learning techniques like convolutional neural networks (CNNs) have been widely used in many fields, e.g., object recognition (high-level vision problem)~\cite{krizhevsky2012imagenet} and image super-resolution or denoising (low-level vision problem)~\cite{ledig2016photo,Chen2016Deep}. 
Recently, the generative adversarial network (GAN) proposed in~\cite{goodfellow2014generative} provides us with a new learning strategy to learn generative neural networks. 
It achieves amazing performance on image generation, which has led an explosion of image-related applications. 
For example, a conditional GAN-based end-to-end image translation method is achieved in~\cite{isola2016image}, which shows excellent performance on different image translation tasks. 
Although many GAN-based methods have achieved encouraging results for image stylization and translation, to our surprise, few of them consider to learn an inverse tone mapping network based on the scheme of GAN. 
Since inverse tone mapping is essentially an image translation operation, the neural network-based model and its generative adversarial learning method should be suitable for our task. 
Our work actually fills the gap between GAN and the application of iTM.

\section{Proposed Methods}

%

We propose an inverse tone mapping network (iTMN) that is able to generate HDR images with satisfying visual effects and robust to different objects and scenes. 
We learn the proposed network with a generative adversarial regularizer. 
In particular, the proposed network is a generator that takes one LDR image in RGB channel as input and produces one corresponding HDR image as output. 
When learning the generator, we further come up with a discriminator that aims to distinguish the real HDR images from the ``fake'' ones produced by the generator. 
Learning the generator and the discriminator jointly corresponds to a GAN-based learning strategy, in which besides the content-based loss a generative adversarial regularizer measuring the loss between the distribution of real data and that of ``fake'' data is considered. 

Denote the batch of LDR images and the corresponding HDR images from the training set as $L$ and $H$. 
The optimization problem corresponding to learning our iTMN is
\begin{eqnarray}\label{opt}
\begin{aligned}
\min_{G}\max_{D}~\underbrace{\lambda\mathcal{L}_{G}(L, H)}_{\text{content loss}} + \underbrace{\mathcal{R}_{G,D}(L, H)}_{\text{adversarial regularizer}}.
\end{aligned}
\end{eqnarray}
Here $G$ and $D$ represent the generator and the discriminator we want to learn, whose architectures are given in Fig.~\ref{fig:1}. 
$\mathcal{L}$ and $\mathcal{R}$ represent the loss function and the regularizer in the objective function, whose importance is controlled by $\lambda$.
In particular, our generator is built with the U-Net in~\cite{ronneberger2015u}, which is an encoder-decoder network with skip connections. 
Each layer is a convolution layer with batch-normalization layer and LeakyReLU as activation function, but sigmoid as activation function at the last layer. 
Our architecture consists of 10 layers with 5 convolution layers and 5 conv-transpose (or called deconvolution) layers. 
Following the guidelines summarized by~\cite{arjovsky2017wasserstein}, we build the architecture using LeakyReLU and avoiding max-pooling layer. 
U-Net is proved to perform well in multi-scale tasks~\cite{Xu2017Two}. 
Considering the requirement that the proposed iTM operation should be multi-scale invariant, we think U-Net should be suitable for our work. 
For our discriminator, its convolution layer is a convolution-BatchNorm-LeakyReLU module and its fully-connected layers take LeakyReLU as activation function.

$\mathcal{L}_G(L, H)$ represents the content-related loss. 
In our work, we propose a hybrid content loss considering the mean squared error between the generated HDR images and the real ones (MSE) and that between their differential results (dMSE) jointly. 
In particular, $\mathcal{L}_G(L, H)$ can be rewritten as
\begin{eqnarray}\label{contentloss}
\begin{aligned}
&\mathbb{E}_{(L,H)\sim p_{data}}[\underbrace{\|G(L)-H\|_F^2}_{\text{MSE loss}} \\
&~~+ \underbrace{\alpha(\|d_xG(L)-d_xH\|_F^2+\|d_yG(L)-d_yH\|_F^2)}_{\text{dMSE loss}}],
\end{aligned}
\end{eqnarray}
where $\mathbb{E}[\cdot]$ calculates the expectation of input, and $(L,H)\sim p_{data}$ means sampling pairs $(L,H)$ from the training set. 
$G(L)$ is the generated HDR image.
$\|\cdot\|_F$ is the Frobenius norm of tensor, and $d_x$ and $d_y$ calculates the horizontal and the vertical differential for each channel of image. 
The first MSE loss is a pixel-level loss that is widely used in many image processing tasks~\cite{ledig2016photo}. 
Moreover, we use the second dMSE loss to further evaluate the difference between the real and ``fake'' HDR images in a deeper level.
Introducing the dMSE loss helps us to suppress the problem of over-smoothness.

$\mathcal{R}_{G,D}(L,H)$ is the proposed generative adversarial regularizer, which is borrowed from the definition of GAN~\cite{goodfellow2014generative}. 
This regularizer is related to both the generator and the discriminator, which has a particular form as
\begin{eqnarray}\label{advreg}
\begin{aligned}
\mathbb{E}_{H\sim p_{data}}[\log(1-D(H))] + \mathbb{E}_{L\sim p_{data}}[\log(D(G(L)))].
\end{aligned}
\end{eqnarray}
Here $H\sim p_{data}$ ($L\sim p_{data}$) means sampling HDR images (LDR images) from the training set.

The min-max problem (\ref{opt}) is a game for getting a better generated image and distinguishing the ``real'' from ``fake'' with higher accuracy, which require us to updating $G$ and $D$ alternatively. 
Specifically, we decompose (\ref{opt}) into the following two problems and solve them iteratively. 
Denote the initial generator and discriminator in the $k$-th iteration as $G_k$ and $D_k$, respectively. 
We have
\begin{eqnarray}\label{minmax}
\begin{aligned}
&{G}_{k+1}=\arg\sideset{}{_{G}}\min~\lambda\mathcal{L}_{G}(L, H) + \mathcal{R}_{G, D_{k}}(L, H),\\
&{D}_{k+1}=\arg\sideset{}{_{D}}\max~\mathcal{R}_{G_{k+1}, D}(L, H).
\end{aligned}
\end{eqnarray}
On the one hand, when learning the generator we encourage our network to favor solutions. 
By minimizing the content loss, the generated HDR image should approach to the ground truth in pixel level. 
By minimizing the regularizer, the generated HDR images try to fool the discriminator and its distribution should be close to that of real HDR images. 
On the other hand, when optimizing the discriminator, we enhance the discriminator to distinguish the difference between the real HDR images and the generated ones, such that the generator should be further optimized in the next iteration.

Such an algorithmic framework can be viewed as an imitation of human-based image editing process. 
Learning generator is similar to trying and fine-tuning different maps while learning discriminator is similar to comparing the generated image with the experienced samples in our minds. 
Finally, an ``experienced editor'' (i.e., a good generator) are trained under strict comparison criteria (i.e., a good discriminator).

\section{Experiments}
As we concerned, there are no related dataset suitable for inverse tone mapping and few corresponding HDR and LDR images.
We build our own training dataset with 2660 HDR images, which are converted to RGB channel and resized in $512 \times 512$.
The pixel values originally quantified in 10-bit comply with the color gamut in BT.2020 standard.
To construct corresponding LDR images, we apply tone mapping operators including ReinhardTMO~\cite{reinhard2002photographic} and so on, and all the pixels are normalized into $[0,1]$. 
When training the architecture, we randomly choose a batch of $6$ images in each iteration and train the generator and discriminator for $80000$ iterations.
In each iteration, we train our generator and discriminator via alternatively optimizing (\ref{minmax}).
The optimizer is RMSProp with a learning rate of step decline from $10^{-4}$.
The parameter $\lambda$ is set to $10^{4}$ and $\alpha$ is set to $10^{5}$.

Evaluation metrics for HDR contents are different from common LDR image processing.
For iTM methods, the most common evaluation metric is HDR-VDP-2~\cite{mantiuk2011hdr}, which compares the test image with a reference image and predicts quality scores expressing the quality degradation.
HDR-VDP-2 can be expressed as a mean-opinion-score to evaluate the quality of the reconstructed HDR images intuitively.
In addition, we apply mPSNR in pixel-wise image quality and SSIM in structural image similarity, to compare various methods.

\subsection{Comparison with Baselines}
\begin{figure*}[!t]
\centering
\includegraphics[width=1\linewidth]{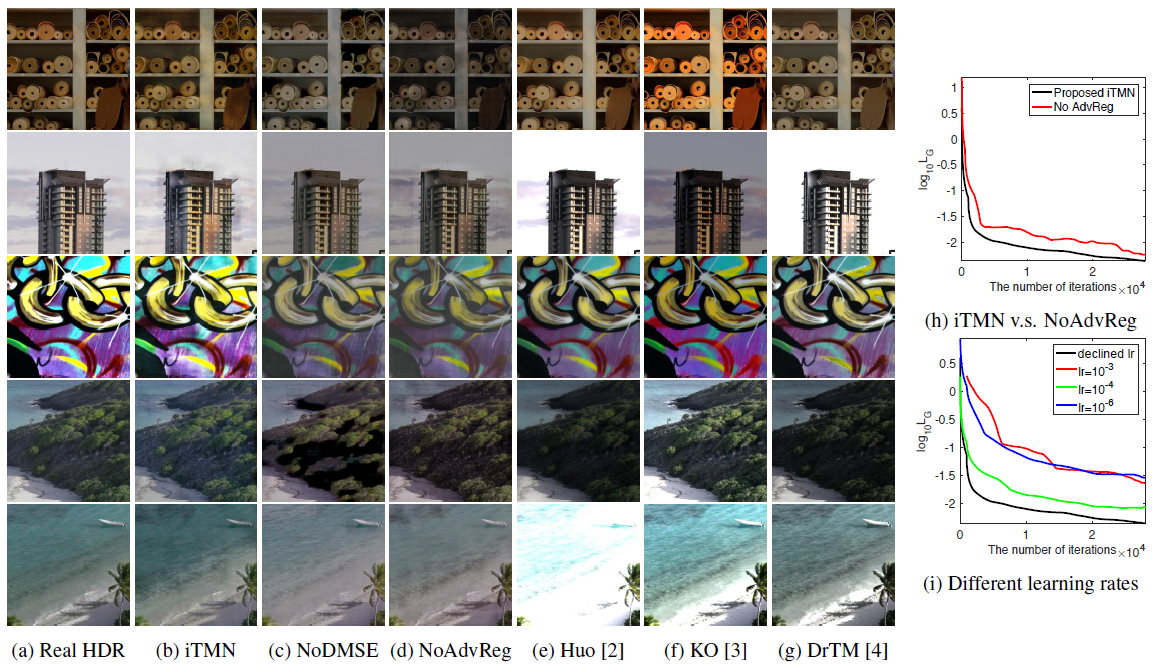}
\vspace{-5pt}
\caption{\small (a-g) Comparisons for various methods. (h, i) Comparisons on convergence.
}
\label{fig:2}
\end{figure*}

To demonstrate the superiority of our method (\textbf{iTMN}), we test it on a large dataset and compare it with existing state-of-the-art methods.
Specifically, the competitors include the methods respectively proposed by \textbf{Huo}~\cite{huo2014physiological}, Kovaleski~\cite{kovaleski2014high} (named as \textbf{KO}) and Endo~\cite{Endo2017Deep}(named as \textbf{DrTM}). 
Additionally, to prove the necessity of the proposed dMSE loss and the generative adversarial regularizer, we propose two variants of our iTMN: the iTMN without dMSE loss (\textbf{NoDMSE}) and the iTMN without adversarial regularizer (\textbf{NoAdvReg}). 
The luminance of the competitors are set to 1000nits as the original HDR, and other parameters are set as the corresponding papers did. 
The numerical comparisons for various methods on different evaluation metrics are listed in Table~\ref{tab:1}, and some visual comparisons are displayed in Fig~\ref{fig:2} (a-g).

\begin{table}[!t]
\centering
\small
\caption{Comparison for various methods}
\vspace{-5pt}
\begin{tabular}{@{\hspace{1pt}}c@{\hspace{1pt}}
|@{\hspace{1pt}}c@{\hspace{1pt}}
@{\hspace{2pt}}c@{\hspace{1pt}}
@{\hspace{1pt}}c@{\hspace{1pt}}
@{\hspace{2pt}}c@{\hspace{2pt}}
@{\hspace{2pt}}c@{\hspace{2pt}}
@{\hspace{2pt}}c@{\hspace{1pt}}}
	\hline\hline
    Method &\footnotesize iTMN &\footnotesize NoDMSE &\footnotesize NoAdvReg &\footnotesize Huo~\cite{huo2014physiological} &\footnotesize KO~\cite{kovaleski2014high} &\footnotesize DrTM~\cite{Endo2017Deep} \\
	\hline
	HDR-VDP-2 &\textbf{73.77} &72.47 &70.52 &60.84 &56.45 &73.21 \\
	mPSNR     &\textbf{21.61} &20.62 &20.07 &20.64 &14.25 &18.73\\
	SSIM      &\textbf{0.8098} &0.7298 &0.7242 &0.7707 &0.6459 &0.7968\\
	\hline\hline
\end{tabular}
\label{tab:1}
\end{table}

We can find in Table~\ref{tab:1} that our method obtains superior results (i.e., higher HDR-VDP score, mPSNR and SSIM) to its competitors consistently. 
Higher HDR-VDP-2 score means that the HDR images obtained by our method has less degradation compared to the ground truth. 
Higher SSIM means our method performs better in image structural quality, and larger mPSNR imples less distortion in pixel-level. 
The visualizations of samples in our dataset in Fig~\ref{fig:2} (a-g) further verify our claim. 
In particular, the HDR images obtained by our iTMN are very close to the real HDR images, while the results corresponding to Huo, KO, and DrTM are unstable, which suffer to serious contrast and color distortions.
More HDR results can be found on our website\footnote{\url{http://medialab.sjtu.edu.cn/projects/SDR2HDR/itmn.html}}.
These results imple that with the help of an explicit objective function based on the ground truth, the learning-based approach can outperform traditional operators distinctly.

For our iTMN, besides the MSE loss, the dMSE loss reduces the difference between the real HDR images and the estimation results in the field of gradient, and the generative adversarial regularizer further imposes constraints on the difference in a higher and more abstract level. 
Both of them provide our problem and the corresponding training process with useful constraints. 
The usefulness of these two components is proven --- in Table~\ref{tab:1} and Fig~\ref{fig:2} (a-g), the results of NoDMSE and NoAdvReg are worse than those of our iTMN. 
Additionally, adding the generative adversarial regularizer helps us to improve the stability and the convergence of loss function $\mathcal{L}_G$, which is verified in Fig~\ref{fig:2} (h). 
With the increase of the number of iterations, the loss function corresponding to our iTMN reduces consistently and converges more quickly than that of NoAdvReg.
In a limited number of iterations, our iTMN reach a better performance in loss descent.

\subsection{Parameter Sensitivity}

We validate the robustness of our iTMN to its parameters, including learning rate, the $\alpha$ in~(\ref{contentloss}) and the $\lambda$ in~(\ref{minmax}). 
When analyzing learning rate, we apply different learning rates and show their influences on the convergence of loss function $\mathcal{L}_G$. 
The results are shown in Fig.~\ref{fig:2} (i). 
We can find that the step-declining learning rate beginning with $10^{-4}$ achieves the best convergence. 
When the learning rate is too large (i.e., $10^{-3}$) or too small (i.e., $10^{-6}$), the loss function converges much more slowly.

\begin{table}[!t]
\small
\centering
\caption{Comparison on different parameter values}
\vspace{-5pt}
\begin{tabular}{@{\hspace{2pt}}c@{\hspace{2pt}}|
@{\hspace{2pt}}c@{\hspace{2pt}}
@{\hspace{2pt}}c@{\hspace{2pt}}
@{\hspace{2pt}}c@{\hspace{2pt}}
@{\hspace{2pt}}c@{\hspace{2pt}}
@{\hspace{2pt}}c@{\hspace{2pt}}}
\hline\hline
$\alpha$/$\lambda$    &  $10^{3}$/$10^{4}$  &  $10^{5}$/$10^{4}$  &  $10^{7}$/$10^{4}$  &  $10^{5}$/$10^{2}$ &  $10^{5}$/$10^{6}$\\
\hline
HDR-VDP-2 &  71.21   &  \textbf{73.77}   &  72.80   &  72.74   &    73.00\\
mPSNR     &  21.20   &  \textbf{21.61}   &  18.79   &  18.85   &    18.78\\
SSIM      &  0.7842  &  \textbf{0.8098}  &  0.6927  &  0.7634  &   0.7210\\
\hline\hline
\end{tabular}
\label{tab:2}
\end{table}

The importance of the dMSE loss and that of the content loss in generator are controlled by $\alpha$ and $\lambda$, respectively. 
We investigate the proper values of these two parameters and Table~\ref{tab:2} compares the performance of our iTMN under different configurations. 
In the case that $\lambda=10^4$ and $\alpha=10^5$, the best performance is achieved in relative to other combinations of parameters. 
Note that when $\alpha<10^3$ or $\lambda>10^6$, the value of the dMSE loss or that of the adversarial regularizer will be ignorable compared to the MSE loss and our iTMN will degrades to the NoDMSE or the NoAdvReg method.

\section{Conclusions and Future Work}
We have presented a novel inverse tone mapping network trained based on a generative adversarial regularizer.
Our method learns an end-to-end mapping from LDR to HDR. 
The superiority of our method to others shows the potentials of learning-based method to the application of iTM.
In the future, we plan to further extend our method to video HDR processing.

\section{Acknowledgement}
This work was supported by NSFC (61671296 and 61521062) and the Shanghai Key Laboratory of Digital Media Processing and Transmissions.

\clearpage
{
\bibliographystyle{IEEEbib}
\bibliography{citation}
}

\end{document}